\documentclass[aps,prb]{revtex4}
\usepackage{psfrag,graphicx}

\begin{document}

\title{Quantum Description of Shuttle Instability\\
in Nanoelectromechanical Single Electron Transistor}

\author{D. Fedorets}
\email[]{dima@fy.chalmers.se}

\affiliation{Department of Applied Physics, Chalmers University of
  Technology \\ and G\"oteborg University, SE-412 96 G\"oteborg,
  Sweden}

\date{\today}
\begin{abstract}
The voltage dependence of nanoelectromechanical effects in a
system where the quantized mechanical vibrations of a quantum dot
are coupled to coherent tunneling of electrons through a single
level in the dot is studied.
 It is found that there are two different regimes depending on the
 value of an applied voltage.
If the bias voltage is below a certain threshold value,
 then the state of the mechanical subsystem is
located near its ground state.
If the bias voltage is above the  threshold value
 then the system becomes  unstable which manifests itself in
 the expectation value of the displacement being
an oscillating function of time with an exponentially increasing
amplitude. This can be interpreted as a shuttle instability in a
quantum regime.
\end{abstract}


\maketitle



Nanoelectromechanical systems (NEMS), where  electronic and
mechanical degrees of  freedom are coupled is a new and fast
growing branch of condensed matter physics
\cite{Roukes00,Craighead00}. One such a system, a single-electron
transistor (SET) where the conducting island has a vibrational
degree of freedom associated with its center-of-mass motion
(Fig.1), a so-called nanoelectromechanical-SET (NEM-SET), has been
attracting a great deal of attention recently, both theoretically
\cite{Gorelik98,Nishiguchi01,Boese00,
Fedorets02a,McCarthy02,Nord02,Armour02,Alexandrov02,Fedorets02b,Novotny03,
Mitra03,Aji03,Lu03,Flensberg03a, Flensberg03b}
 and experimentally \cite{Park00, Erbe01}.
\begin{figure}[htbp]
  \begin{center}
    \psfrag{x}{x}
    \psfrag{+eV}{$\mu_L=\frac{eV}{2}$}
    \psfrag{0}{$\mu_R=-\frac{eV}{2}$}
    \psfrag{Lead}{Lead}
    \psfrag{Dot}{Island}
    \includegraphics[width = 7cm]{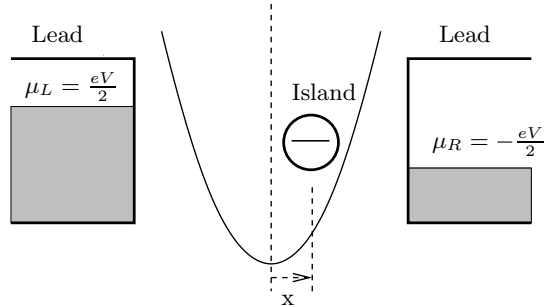}
    \vspace*{0.3cm}
    \caption{Model shuttle system consisting of a movable  conducting
    island placed
    between two leads.
    An effective elastic force acting on the dot from the leads is
    described by the parabolic potential.
      }\label{system}
  \end{center}
\end{figure}

In Ref.~\onlinecite{Gorelik98} it was shown, that the metallic
grain placed between the two leads of NEM-SET becomes unstable and
the periodic mechanical motion of the grain develops
 if a large bias voltage is applied between the
leads. This phenomenon is usually referred to as a shuttle
instability (see the review Ref.~\onlinecite{Shekhter02}). A theory
of the shuttle instability developed in
Ref.~\onlinecite{Gorelik98} was based on the assumptions that both
the charge on the island and
 its trajectory are well defined quantities.

When we decrease the island size two different quantum effects
manifest themself. Firstly, the electron energy level spacing in a
nanometer-size grain is of the order of $10$~K and resonant
tunneling effects become essential at small enough temperatures.
Therefore the single-electron energy spectrum can not be treated
as continuous any more as it was done in
Ref.~\onlinecite{Gorelik98}. In this case the characteristic de
Broglie wave length associated with the island can still be much
shorter than the length scale of the spatial variations of the
"mechanical" potential. If so, the motion of the island can be
treated classically. This regime has been studied theoretically in
Ref.~\onlinecite{Fedorets02a}.

Diminishing the size of the island further (down to $7$~\r{A} in
diameter, for $C_{60}$ molecule)
 results in the
quantization of the mechanical motion of the island. A NEM-SET
system in the regime of quantized mechanical motion of the central
island was studied theoretically in Ref.~\onlinecite{Boese00}.
 It was assumed that the phase breaking
processes are strong enough to make the density matrix  diagonal
in the representation of the eigenstates of the quantum oscillator
Hamiltonian $|n\rangle $, which describes the mechanical
subsystem. At the same time it is well known that the expectation
value of the displacement operator in the eigenstates of a quantum
oscillator are zero while it is the coherent state, which is a
coherent superposition of all $|n\rangle $, that approaches a
description of the classical periodic motion of the oscillator as
$\hbar \rightarrow 0$.
 The interesting question arises, therefore,  how the
expectation value of displacement operator evolves in time and
what state results when the formal condition of the shuttle
instability are satisfied for the quantum NEM-SET.

The approach used in Ref.~\onlinecite{Boese00} was further
extended in Ref.~\onlinecite{McCarthy02} to include the position
dependence of the tunneling matrix elements. However, the possibility of
the shuttle instability effects was excluded by assuming a strong
external dissipation. The case of the weak dissipation was
studied in Ref.~\onlinecite{Fedorets02a, Novotny03} in the
limit of large voltages. The effect of a quantized
vibrational mode on electron tunneling through a similar structure
(a chain of three quantum dots) has been studied in
Ref.~\onlinecite{Armour02}.

In this article we will show that if a bias voltage between the
leads is higher then a certain value, then the quantum state
of the central island of the NEM-SET evolves in such a way that
the expectation value oscillates in time with an increasing
amplitude. This results shows that the shuttle instability is a
fundamental phenomenon which exists even when the trajectory of
the island and the charge on it are no longer well-defined.


We use the following Hamiltonian to model our system
\begin{eqnarray}
H = \sum_{\alpha, k} \epsilon_{\alpha k} a_{\alpha k}^{\dag}
a_{\alpha k} +  \epsilon_d(X) c^{\dag} c + \frac{P^2}{2M} +
\frac{Mw_0^2X^2}{2} + \sum_{\alpha} T_{\alpha}(X)
(A_{\alpha}^{\dag} c + c^{\dag} A_{\alpha })\,.
\end{eqnarray}
Here $A_\alpha \equiv \sum_k a_{\alpha k}$,
\begin{eqnarray} \label{eps}
\epsilon_d(X) = \epsilon_0 - d X\,,
\end{eqnarray}
where $d$ is proportional to an electric field
between the leads,
\begin{eqnarray} \label{T}
T_{L,R}(X)=T_0 \exp\{\mp X/\lambda\},
\end{eqnarray}
where $\lambda$ is a characteristic tunneling length. The first
term in the Hamiltonian describes the electrons in the leads, the
second term relates to the single energy level in the central
island, the third and forth terms to the quantized vibrational
degree of freedom associated with center-of-mass motion of the
central island and the last term describes tunneling between the
electrodes and the island. All energies are measured from the
Fermi energy of the leads. We assume that the electrons in each
electrode are non-interacting with a constant density of states
and that all relevant energies are small compared to the level
spacing in the central island which for typical systems under
consideration exceeds $100$~meV. In this case only one single
level in the central island is relevant to the problem.

It is convenient to introduce dimensionless operators for
displacement, $x\equiv X/r_0$, and momentum, $p\equiv r_0
P/\hbar$, where $r_0 \equiv \sqrt{\hbar/(Mw_0)}$, and then measure
all lengths in units of $r_0$ and all energies in units of $\hbar
w_0$.

The Hamiltonian can now be written as
\begin{equation}\label{}
    {H} = H_{el} + H_{dot} + H_T\,,
\end{equation}
where
\begin{eqnarray}
H_{dot} &\equiv& \frac{1}{2}\left[{p^2} + {x^2}\right] +
 \epsilon_d(x) c^{\dag} c \,,\\
H_{el} &\equiv& \sum_{\alpha, k} \epsilon_{\alpha k} a_{\alpha
k}^{\dag}
a_{\alpha k}\,,\\
H_T &\equiv& \sum_{\alpha} \left[ A_{\alpha}^{\dag} C_\alpha +
 C_\alpha^{\dag} A_{\alpha} \right] \,,\quad C_\alpha \equiv T_\alpha(x)\, c\,.
\end{eqnarray}

We will use an equation of motion (EOM) approach to study the time
evolution of the operator $x$ in the Heisenberg picture. In this
approach we deal with the algebra of the time-dependent operators
$a_{\alpha k}(t)$, $c(t)$, $x(t)$, $p(t)$ in the Heisenberg
representation whose evolution can be described by a coupled
system of EOMs. The EOM for the operator $a_{\alpha k}(t)$ can be
formally solved and  eliminated from the system. The reduced
system of EOMs is local in time under the condition of the wide-band
approximation (${\cal D}_\alpha = {\rm const}$, where ${\cal
D}_\alpha$ is a density of state in the lead $\alpha$). It is
shown in Appendix \ref{appendixA} that the evolution of an
operator $S=S(x,p, c^{\dag} c)$ is described by the following
quantum Langevin equation
\begin{eqnarray}\label{eqMotionDot_2}
\dot{S} = - i[S, H_{dot}] - \sum_{\alpha}\left\{
i\sqrt{\eta_\alpha}
 a_{\alpha}^{\dag} - \frac{\eta_\alpha}{2} C_\alpha^{\dag} \right\}
[S,C_\alpha] -  \sum_{\alpha}[S,C_\alpha^{\dag}] \left\{
i\sqrt{\eta_\alpha}a_{\alpha} + \frac{\eta_\alpha}{2}
C_\alpha\right\}\,,
 \end{eqnarray}
 where
\begin{equation}
a_{\alpha}(t) \equiv \frac{1}{\sqrt{\eta_\alpha}} \sum_{k} e^{-i
\epsilon_{\alpha k}t} a_{\alpha k}(0)\,,\quad \eta_\alpha \equiv
2\pi{\cal D}_\alpha\,.
\end{equation}
For complete description we need also to know the evolution of
$c(t)$, which as shown in Appendix \ref{appendixB} is given by
\begin{eqnarray}
c(t) = \sum_{\alpha, k}   e^{-i \epsilon_{\alpha k} t} B_{\alpha
k}(t\,,0)
  a_{\alpha k}(0)
  \,,\label{c_3}
\end{eqnarray}
where
\begin{eqnarray}
B_{\alpha k}(t\,,t')\equiv \frac{1}{i} {\cal T}\int_{t'}^{t}dt_1
 e^{i\int_{t_1}^{t} dt_2 [\epsilon_{\alpha
k}-\hat{\epsilon}_d+i\frac{\hat{\Gamma}}{2}]}
\hat{T}_{\alpha}(t_1)\,,
\end{eqnarray}
${\cal T}$ is time-ordering operator, $\hat{\Gamma}(t) \equiv
\sum_\alpha \eta_\alpha \hat{T}_\alpha^2(t)$, $\hat{T}_\alpha(t)
\equiv {T}_\alpha(x(t))$, $\hat{\epsilon}_d(t) \equiv
{\epsilon}_d(x(t))$ and $x(t)$ is the operator $x$ in the
Heisenberg picture.

We substitute  $x$ and $p$ in Eq.~(\ref{eqMotionDot_2}) and get
\begin{eqnarray}
\dot{x} = p\,,\quad \dot{p} &=& -x + \hat{F}\,,
\end{eqnarray}
or
\begin{eqnarray}
\ddot{x} + x = \hat{F}\,,\label{newton_1}
\end{eqnarray}
 where
\begin{eqnarray}
\hat{F} \equiv dc^{\dag}c + 2\lambda^{-1} \sum_{\alpha}
(-1)^{\alpha}\sqrt{\eta_\alpha}\, \hat{T}_\alpha {\rm Re}
\left\{a_\alpha^{\dag} c\right\}\,.
\end{eqnarray}

To study the shuttle instability in the quantum regime we assume
that dimensionless parameters $1/\lambda$ and $d$ are small which
allows us to linearize the problem with respect to $x$.
Eq.~(\ref{c_3}) then reads
\begin{eqnarray}
c(t)
 = T_0 \sum_{\alpha, k}  e^{-i\epsilon_{\alpha k}t}
\left\{G_{\alpha k}^{+} + i\left[d G_{\alpha k}^{+} +
\frac{(-)^\alpha}{\lambda}\right] X_{\alpha k}(t)\right\}
a_{\alpha k}(0) \,.
\end{eqnarray}
Here $G_{\alpha k}^{+}\equiv [\epsilon_{\alpha
k}-\epsilon_0+i{\Gamma}/{2}]^{-1}$, $\Gamma \equiv \sum_{\alpha}
\eta_\alpha T_0^2$, $T_0$ is defined in Eq.~(\ref{T}),
\begin{eqnarray}
X_{\alpha k}(t) \equiv \int_{0}^{t} d\tau e^{ i[\epsilon_{\alpha
k}-\epsilon_0+i\frac{\Gamma}{2}](t-\tau)}x(\tau)\,
\end{eqnarray}
and the linearized force operator $\hat{F}$ has the following form
\begin{eqnarray}
\hat{F} \equiv   \hat{F}_1 + \hat{F}_2\,,
\end{eqnarray}
where
\begin{eqnarray}
\hat{F}_{1} \equiv   T_0^2\sum_{}  e^{i(\epsilon_{\alpha
k}-\epsilon_{\beta q})t} a_{\alpha k}^{\dag}(0) a_{\beta q}(0)
\left\{dG_{\alpha k}^{-}G_{\beta q}^{+} + \lambda^{-1} \left[
(-)^\alpha G_{\beta q}^{+} + (-)^\beta G_{\alpha k}^{-}\right]
 \right\} \,
\end{eqnarray}
is the term of the first order in the small parameters $d$ and
$\lambda^{-1}$ and
\begin{eqnarray}
\hat{F}_{2} \equiv    T_0^2\sum_{} e^{i(\epsilon_{\alpha
k}-\epsilon_{\beta q})t} a_{\alpha k}^{\dag}(0) a_{\beta q}(0)
\left\{ - \frac{x}{\lambda^{2}} \left[ G_{\alpha k}^{-} + G_{\beta
q}^{+} \right]  + i  \left[ X_{\beta q} - X_{\alpha k}^{*}\right]
\left[ dG_{\alpha k}^{-} + \frac{(-)^\alpha}{\lambda}\right]
\left[ dG_{\beta q}^{+} + \frac{(-)^\beta}{\lambda}\right]
 \right\} \,
\end{eqnarray}
is the second order term in $d$ and $\lambda^{-1}$.
Thus the above linearization of the force $F$ corresponds to its
expansion up to  second order in $d$ and $\lambda^{-1}$.

Averaging Eq.~(\ref{newton_1}) with respect to the initial density
operator we obtain EOM for the expectation value of the
displacement of the dot. If we only take into account effects up
to second order in $T_0$ (Born approximation), we can make the
following factorization in the RHS of Eq.~(\ref{newton_1})
\begin{equation}
    \left<x(t)a_{\alpha k}^{\dag}(0) a_{\beta q}(0)\right> \approx
    \left<x(t)\right> f_{\alpha k} \delta_{\alpha k,\beta q}\,,
\end{equation}
where  $f_{\alpha k} \equiv [1+e^{(\epsilon_{\alpha
    k}\mp eV/2)/(kT)}]^{-1}$. Then
\begin{eqnarray}
\ddot{\bar{x}} + \bar{x} =
\bar{F}\left[x(\tau)\right]\,,\label{newton_2}
\end{eqnarray}
where $\bar{x}(t) \equiv \left<x(t)\right> -x_0$, $x_0 \equiv
(d/2)\sum_\alpha f_{\alpha}(\epsilon_0)$, $f_\alpha(E) \equiv
[1+e^{(E \mp eV/2)/(kT)}]^{-1}$ and
\begin{eqnarray}
 \bar{F}\left[x(\tau)\right] \equiv   -2 T_0^2 \sum_{}f_{\alpha k}
 {\rm Im} \left[ X_{\alpha k} \right]
  \left|d
G_{\alpha k}^{+} + \frac{(-)^\alpha}{\lambda}\right|^2 \,.
\label{classic_force}
\end{eqnarray}
It is remarkable that Eq.~(\ref{newton_2}) with $\bar{F}$ given by
Eq.~(\ref{classic_force}) is the same as the  Newton's
equation used in Ref.~\onlinecite{Fedorets02a}. It is well-known
that  for an isolated quantum harmonic oscillator the equation of
motion for the expectation values of displacement and momentum is exactly the
same as the classical Hamiltonian equations (Ehrenfest theorem). However,
it is not obvious that this
statement is  valid for the harmonic oscillator
coupled to some degrees of freedom. Therefore, it follows from
Eq.~(\ref{classic_force}) that
the Ehrenfest theorem is valid for the system under consideration
 if the tunneling is weak.

We solve Eq.~(\ref{newton_2}) by Laplace transforms and obtain
\begin{eqnarray}
 \bar{x}(t) \approx -{x_0 e^{rt}} \cos t
\,,\label{solution}
\end{eqnarray}
where
\begin{eqnarray}
 r \equiv  \frac{\Gamma}{8}
\sum_{\alpha}\left\{\left[d+\frac{1}{\lambda}\right]^2
 f_{\alpha,+1}-
\left[d-\frac{1}{\lambda}\right]^2 f_{\alpha,-1}
  \right\}\,,
\label{increment}
\end{eqnarray}
 and $f_{\alpha, \pm 1} \equiv
f_{\alpha}(\epsilon_0\pm 1)$. The sign of $r$ depends on the bias
voltage and can be easily analyzed if the temperature is zero. If
the bias voltage is below the threshold value $V_c$, defined by
$eV_c=2(\epsilon_0+1)$, then $f_{\alpha,+1}=0$ and $r\leq0$. In
this case $\bar{x}(t)$ exponentially goes to zero (point
$\bar{x}=0$ is stable). If the voltage is above the threshold
value, then $f_{L,\pm1}=1$, $f_{R,\pm1}=0$ and $r=\Gamma
d/(2\lambda)$. In this case $\bar{x}(t)$ oscillates with
exponentially growing amplitude (point $\bar{x}=0$ is unstable)
and the corresponding state of the mechanical subsystem moves
further and further from the ground state.

 In conclusion,  we have studied  stability of a system where the quantized
mechanical vibrations of a quantum dot are coupled to coherent
tunneling of electrons through a single level in the dot. Two
different regimes have been found. For bias voltages above a
certain threshold value (which, in the case of zero temperature,
is defined by the position of the level in the dot plus the energy
of a vibrational quantum) the system is unstable. In this regime
the expectation value of the displacement is an oscillating
function of time with an exponentially increasing amplitude, which
is the signature of a shuttle instability in a quantum regime.
Below the threshold the system is stable even without any external
damping of the dot's motion.

The author would like to thank L. Y. Gorelik and R. I. Shekhter for
valuable discussions.

\appendix
\section{}\label{appendixA}
The equation of motion for $a_{\alpha k}(t)$ is given by
\begin{eqnarray}
\partial_t a_{\alpha k} = i[H, a_{\alpha k}]
 = -i\epsilon_{\alpha k} a_{\alpha k} - iC_\alpha \,.
 \label{eqMotionLead}
\end{eqnarray}

We formally solve Eq.~(\ref{eqMotionLead}) to obtain
\begin{equation}
a_{\alpha k}(t) = e^{-i \epsilon_{\alpha k}t} a_{\alpha k}(0) - i
\int_{0}^t \, d t_1 e^{-i \epsilon_{\alpha k}(t-t_1)}
C_\alpha(t_1)\,. \label{a}
\end{equation}
Then by using  the wide-band approximation (${\cal D}_\alpha =
{\rm const}$) we get
\begin{equation}
A_{\alpha}(t) = \sqrt{\eta_\alpha}a_{\alpha}(t) -
 i\frac{\eta_\alpha}{2} C_\alpha(t)\,,
\label{A}
\end{equation}
where
\begin{equation}
a_{\alpha}(t) \equiv \frac{1}{\sqrt{\eta_\alpha}} \sum_{k} e^{-i
\epsilon_{\alpha k}t} a_{\alpha k}(0)\,,\quad \eta_\alpha \equiv
2\pi{\cal D}_\alpha\,.
\end{equation}

If $S=S(x,p, c^{\dag} c)$, then its equation of motion is
\begin{eqnarray}\label{eqMotionDot_1}
\dot{S} = -i[S, H_{dot}] - i\sum_{\alpha}\left\{ A_{\alpha}^{\dag}
[S,C_\alpha] + [S,C_\alpha^{\dag}]A_{\alpha} \right\}\,.
 \end{eqnarray}

Combining Eqs.~(\ref{A}) and (\ref{eqMotionDot_1}) yields
Eq.~(\ref{eqMotionDot_2}).

\section{}\label{appendixB}
The equation of motion for $c(t)$ reads
\begin{eqnarray}
\dot{c}(t) = i[H, c]= -i\hat{\epsilon}_d(t) c(t) - i\sum_{\alpha}
\hat{T}_\alpha(t) A_{\alpha}(t)\,, \label{c_1}
\end{eqnarray}
where $\hat{\epsilon}_d(t) \equiv {\epsilon}_d(x(t))$,
$\hat{T}_\alpha(t) \equiv {T}_\alpha(x(t))$ and $x(t)$ is
operator $x$ in the Heisenberg picture. Combining Eqs.~(\ref{A})
and (\ref{c_1}) gives
\begin{eqnarray}
\dot{c}(t) =
-i\left[\hat{\epsilon}_d(t)-i\frac{\hat{\Gamma}(t)}{2}\right] c(t)
- i\sum_{\alpha} \sqrt{\eta_\alpha} \,\hat{T}_\alpha(t)
a_{\alpha}(t)\,.\label{c_2}
\end{eqnarray}
The solution of Eq.~(\ref{c_2}) for $t \gg \,\parallel
\hat{\Gamma}
\parallel^{-1}$ is given by Eq.~(\ref{c_3}).


\end{document}